\documentclass[showpacs,prl,twocolumn,amsmath,amssymb]{revtex4-1}
\usepackage{graphicx}
\usepackage{amsmath,amsfonts}
\usepackage{dcolumn}
\usepackage{bm}
\usepackage{slashed}
\usepackage{nicefrac}
\usepackage{epstopdf}
\usepackage{color}
\def\bra{\langle}
\def\ket{\rangle}

\newcommand{\R}{\mathbb{R}}
\newcommand{\cD}{\mathcal{D}}

\newcommand{\cH}{\mathcal{H}}
\newcommand{\cN}{\mathcal{N}}
\def\dk#1#2{\frac{ d^{#2}{#1} }{ (2\pi)^{#2} }} 
\def\Ei{\mathrm{Ei}}
\begin{document}

\title{Unifying renormalization group and the continuous wavelet transform}
\author{M. V. Altaisky}
\affiliation{Space Research Institute RAS, Profsoyuznaya 84/32, Moscow, 117997, Russia}
\email{altaisky@mx.iki.rssi.ru}

\date{May 26, 2016} 
    
\begin{abstract}
It is shown that the renormalization group turns to be a symmetry group in a theory 
initially formulated in a space of scale-dependent functions, i.e, those 
depending on both the position $x$ and the resolution $a$. Such theory, earlier 
described in \cite{Altaisky2010PRD,AK2013}, is finite by construction. The space of 
scale-dependent functions $\{ \phi_a(x) \}$ is more relevant to physical reality than 
the space of square-integrable functions $\mathrm{L}^2(\R^d)$, because due to the 
Heisenberg uncertainty principle, what is really measured in any experiment is 
always defined in a region rather than point. The effective action $\Gamma_{(A)}$ of our theory turns to be complementary to 
the exact renormalization group effective action. The role of the regulator is played 
by the basic wavelet -- an ''aperture function'' of a measuring device used to produce 
the snapshot of a field $\phi$ at the point $x$ with the resolution $a$. The standard RG 
results for $\phi^4$ model are reproduced.     
\end{abstract}

\maketitle
\section{Introduction}
Renormalization group (RG) was discovered by Stueckelberg and Petermann as a group 
of infinitesimal reparametrizations of the $S$ matrix emerging after the cancellation of 
ultraviolet divegences \cite{SP1953}. The RG method has become known in quantum electrodynamics (QED) since Gell-Mann and Low, using the functional equation for the 
renormalized photon propagator, have shown the charge distribution surrounding a 
test charge in vacuum does not at small distances depend on a coupling constant, except for a scale factor, {\em i.e.,} possesses a kind of self-similarity, that enables to 
express a ''bare'' charge at small scale using the measured value at large scale \cite{GL1954}. This result was latter generalized for massive propagators and 
vertices in QED using the theory of the Lie groups, from where the term {\em renormalization group} comes \cite{BSh1956}. The latter result does not refer to 
the existence of the  ''bare'' charges and is just a set of group functional 
relations between the vertices and the propagators at different scales.

The RG method has become popular in utmost all branches of science when entered both the 
statistical physics and the field theory as a method of treating fluctuations of different scales not all at once, but successively scale by scale \cite{Wilson1971a,WK1974}. This resulted in an elegant theory of critical phenomena and 
was later generalized to many other stochastic systems \cite{Vasiliev2004}. 

Interestingly, the same idea of separating the fluctuations of different scales has been 
implemented, basically in experimental data processing, in a quite different way: using 
wavelets. This was first done in geophysics \cite{GGM1984,GM1984}, and then spread over 
all possible data, from face recognition and medical imaging to high energy physics and cosmology \cite{Daub10}. The only intersection of the RG and the wavelet method seems 
to be the the lattice regularization in quantum field theory (QFT), which can be performed either by standard lattice methods, or by using the the discrete wavelet basis \cite{Caroll1993,Battle1999,Best2000}. 
The connections between these two seemingly different methods are still missing. 

The present paper is an endeavor   to unify the RG method with the continuous 
wavelet transform. To make the consideration simple a theory of a massive scalar field in Euclidean space is considered. 

To illustrate the method we start with Euclidean scalar field theory with $\phi^4$ interaction in $\R^d$ defined by the generating functional 
\begin{align}
Z[J] &=& \cN \int \exp\left(-S_E[\phi] + \int J(x)\phi(x) d^dx\right), \label{gf}\\
\nonumber \hbox{where\ } & &S_E[\phi] = \int d^dx \left[ \frac{1}{2} (\partial\phi)^2 + \frac{m^2}{2} \phi^2  
+ \frac{\lambda}{4!}\phi^4\right] 
\end{align}
is the Euclidean action of the theory, $\cN$ is a formal normalization constant. 
The connected Green functions are given 
by functional derivatives:
\begin{equation}
G^{(n)} = 
\left. { \frac{\delta^n W[J]}{\delta J(x_1) \ldots \delta J(x_n)}
}\right|_{J=0} \label{cgf}, 
\end{equation} 
where $W[J] = \ln Z[J]$ is the connected GF generating functional.


The {\sl effective action functional} is defined via the Legendre transform of $W[J]$:
\begin{equation}
\Gamma[\phi] =   -W[J] + \int J(x) \phi(x) d^dx .
\end{equation} 
The functional derivatives of the effective $\Gamma[\phi]$ are the vertex functions $\Gamma^{(n)}$.
For the above considered scalar field theory \eqref{gf} the (renormalized) vertex function $\Gamma^{(4)}$ accounts for the 
value of the coupling constant at certain (IR) scale  $\lambda=\lambda_{\rm Phys}$. On the other hand, the perturbation expansion of the Green functions starts with the bare value of the coupling constant $\lambda\!=\!\lambda_0$, 
which corresponds to the interaction strength on certain (UV) scale to be dressed by interactions. If the theory is subjected to the UV cutoff $\Lambda$, in  one loop 
approximation,  we get 
\begin{equation}
\Gamma^{(4)} = \lambda_0 - \frac{3}{2}\lambda_0^2 \int_\Lambda
\frac{d^dq}{(2\pi)^d}\frac{1}{q^2+m_0^2} + O(\lambda_0^4), \label{g4uv}
\end{equation}
so that the inversion of this formula with respect to $\lambda=\lambda_{\rm Phys}$ yields the increase of the interaction 
strength for the small scales:  
$
\lambda_0 = \lambda + \frac{3}{2}\lambda^2 \int_\Lambda \ldots .
$
Similar results are obtained with dimensional regularization scheme \cite{Wilson1973}.

In the next two sections I'll remind the construction of the scale-dependent field theory 
using wavelets and present the scale dependence of the vertex functions in one loop 
approximation. The differences between the multiscale theory and the standard one are 
outlined in Conclusions.

\section{Field theory in wavelet representation}
Continuous wavelet transform is a  generalization of the Fourier transform 
for the case when the scaling properties of the theory are important. 
Referring the reader to general reviews on wavelet transform \cite{Daub10,Chui1992}, and 
to the original papers devoted to the application of wavelet transform to quantum 
field theory \cite{Altaisky2010PRD,AK2013,BP2013}, below we remind the basic definitions 
of the wavelet formalism of quantum field theory. 

Let $\cH$ be a Hilbert space of states for a quantum field $|\phi\ket$. 
Let $G$ be a locally compact Lie group acting transitively on $\cH$, 
with $d\mu(\nu),\nu\in G$ being a left-invariant measure on $G$.  
Similarly to the Fourier representation 
$
|\phi\ket=\int |p\ket dp \bra p |\phi\ket,$
any $|\phi\ket \in \cH$ can be decomposed with respect to 
a representation $U(\nu)$ of $G$ in $\cH$ \cite{Carey1976,DM1976}:
\begin{equation}
|\phi\ket= \frac{1}{C_g}\int_G U(\nu)|g\ket d\mu(\nu)\bra g|U^*(\nu)|\phi\ket, \label{gwl} 
\end{equation} 
where $|g\ket \in \cH$ is referred to as an admissible vector, 
or a {\em basic wavelet}, satisfying the admissibility condition 
$$
C_g = \frac{1}{\| g \|^2} \int_G |\bra g| U(\nu)|g \ket |^2 
d\mu(\nu)
<\infty. 
$$
The coefficients $\bra g|U^*(\nu)|\phi\ket$ are referred to as 
wavelet coefficients. 
If the group $G$ is Abelian, the wavelet transform \eqref{gwl} coincides with 
the Fourier transform. 

Next to the Abelian group is the group of the affine transformations 
of the Euclidean space $\R^d$:
\begin{equation}
G: x' = a x + b, x,b \in \R^d, a \in \R_+.  \label{ag1}
\end{equation} 
(For the sake of simplicity we assume the isotropic basic wavelet $g$ and 
drop the $SO(d)$ rotation factor.)
The unitary representation of the affine transform \eqref{ag1} with 
respect to the isotropic basic wavelet $g(x)$ can be written as follows:
\begin{equation}
U(a,b) g(x) = \frac{1}{a^d} g \left(\frac{x-b}{a} \right).
\end{equation}  
(In accordance to previous papers \cite{Altaisky2010PRD,AK2013} we use $L^1$ norm \cite{Chui1992,HM1996} instead of usual $L^2$ to keep the physical dimension 
of wavelet coefficients equal to the dimension of the original fields).

Wavelet coefficients of the Euclidean field $\phi(x)$ with 
respect to the basic wavelet $g(x)$ in  $\R^d$ are  
\begin{equation}
\phi_{a}(b) = \int_{\R^d} \frac{1}{a^d} \overline{g \left(\frac{x-b}{a} \right) }\phi(x) d^dx. \label{dwtrd}
\end{equation} 

The function $\phi(x)$ can be reconstructed from its wavelet coefficients 
\eqref{dwtrd} using the formula \eqref{gwl}:
\begin{equation}
\phi(x) = \frac{1}{C_g} \int \frac{1}{a^d} g\left(\frac{x-b}{a}\right) \phi_{a}(b) \frac{dad^db}{a}.  \label{iwt}
\end{equation}
The normalization 
constant is readily evaluated using Fourier transform:
$
C_g = \int_0^\infty |\tilde g(a)|^2\frac{da}{a}.
$

Substituting \eqref{iwt} into the field theory \eqref{gf} we obtain the 
generating functional for the scale-dependent fields $\phi_a(x)$ 
\begin{widetext}
\begin{align} \nonumber 
Z_W[J_a] &=&\cN \int \cD\phi_a(x) \exp \Bigl[ -\frac{1}{2}\int \phi_{a_1}(x_1) D(a_1,a_2,x_1-x_2) \phi_{a_2}(x_2)
\frac{da_1d^dx_1}{a_1}\frac{da_2d^dx_2}{a_2}  \\
&-&
\int V_{x_1,\ldots,x_4}^{a_1,\ldots,a_4} \phi_{a_1}(x_1)\cdots\phi_{a_4}(x_4)
\frac{da_1 d^dx_1}{a_1} \frac{da_2 d^dx_2}{a_2} \frac{da_3 d^dx_3}{a_3} \frac{da_4 d^dx_4}{a_4} 
+ \int J_a(x)\phi_a(x)\frac{dad^dx}{a}\Bigr], \label{gfw}
\end{align}
\end{widetext}
with $D(a_1,a_2,x_1-x_2)$ and $V_{x_1,\ldots,x_4}^{a_1,\ldots,a_4}$ denoting the wavelet images of the inverse propagator and that of the interaction potential.

Therefore $W_W[J_a] \equiv \ln Z_W[J_a]$ is the generating functional of the connected 
Green functions for the scale-dependent fields $\bra\phi_{a_1}(x_1)\cdots\phi_{a_n}(x_n)\ket_c$. We can define the effective action functional 
\begin{equation}
\Gamma[\phi_a] =   -W_W[J_a] + \int J_a(x) \phi_a(x) \frac{da}{a}d^dx, \label{Gw}
\end{equation} 
which accounts for the 1PI Green functions of the scale-dependent fields; with 
all functional derivatives assumed with respect to the measure $\frac{da}{a}d^dx$.

The Feynman diagram technique for the scale-dependent fields $\phi_a(x)$ is the same 
as for ordinary fields except for \cite{Alt2002G24J,Altaisky2010PRD}:
\begin{itemize}\itemsep=0pt
\item each field $\tilde\phi(k)$ will be substituted by the scale component
$\tilde\phi(k)\to\tilde\phi_a(k) = \overline{\tilde g(ak)}\tilde\phi(k)$.
\item each integration in momentum variable is accompanied by corresponding 
scale integration:
\[
 \dk{k}{d} \to  \dk{k}{d} \frac{da}{a}.
 \]
\item each interaction vertex is substituted by its wavelet transform; 
for the $N$-th power interaction vertex this gives multiplication 
by factor 
$\displaystyle{\prod_{i=1}^N \overline{\tilde g(a_ik_i)}}$.
\end{itemize}
According to these rules, the bare Green function in wavelet representation 
takes the form 
$$
G^{(2)}_0(a_1,a_2,p) = \frac{\tilde g(a_1p)\tilde g(-a_2p)}{p^2+m^2}.
$$ 
The finiteness of the loop integrals is provided by the following rule:
{\em there should be no scales $a_i$ in internal lines smaller than the minimal scale 
of all external lines} \cite{Altaisky2010PRD}. Therefore the integration in $a_i$ variables is performed from 
the minimal scale of all external lines up to the infinity. 
This corresponds to the assumption, that studying a system from outside one should 
not used functions with resolution better than the finest experimentally available 
scale. 
The integration over {\sl all} scales will certainly drive us back to the known divergent theory. 

\section{Renormalization}
Being originated as a method of eliminating divergences, the RG technique has 
further evolved to a functional method, referred to as the {\em functional, or exact, renormalization group} \cite{Polchinski1984,Morris1998,Polonyi2003}, which deal with the so-called {\em effective average action} 
$\Gamma_k$ \cite{Wetterich1991,Wetterich1993}, which accounts for the fluctuations of momenta greater than $k$. In this way the effective action $\Gamma_k$ is a functional 
of fields from which the physical properties at a given scale can be computed. The 
IR limit of the effective action is the exact action, $\Gamma_{k\to0}=\Gamma$;  in the UV limit all fluctuations are suppressed and the 
effective action turns to the bare action $\Gamma_{k\to\Lambda\to\infty} \to S_{\rm bare}$.

Technically the effective action is constructed by the addition of a quadratic 
regularization term $$\Delta S_k[\phi]=\frac{1}{2} \int \frac{d^dq}{(2\pi)^d} \phi(-q)R_k(q)\phi(q)$$ to the Euclidean action of the theory. This leads to the regularized generating 
functional
\begin{equation}
Z_k[j] \equiv e^{W_k[J]} = \int_\Lambda \cD\phi e^{-S[\phi] - \Delta S_k[\phi]+ \int J\phi},
\end{equation}
so that $\Gamma_k[\phi] = -W_k[\phi]-\Delta S_k[\phi] + \int J \phi$.
The main instrument of the exact renormalization group is the flow equation $k\frac{\partial \Gamma_k}{\partial k}=\dots$, which generates the RG flow from large 
to small scales and allows to study the scale behavior of the coupling constant if its value  at some reference scale is known. 

What is proposed in this paper is in some way complementary to the effective action $\Gamma_k$ of the ERG, i.e., it similar to $\Gamma -\Gamma_k$. Indeed, according to the above listed Feynman rules for the multi-scale theory (see \cite{Altaisky2010PRD,AK2013} for details), and to the definition of the effective action 
\eqref{Gw}, we can define the action functional $\Gamma_{(A)}[\phi_a]$, where $A$ is the 
minimal scale of all external lines of the contributing diagrams. This functional 
$\Gamma_{(A)}$ incorporates the scale components of all fields $\phi_a(x)$ with $a\ge A$.

Physically the scale of observation $a$, at which the coupling constant $\lambda(a)$ is called $\lambda_{\rm Phys}$, is not necessarily the $k\to0$ limit: it may be large but finite size of the system. {\sl Vice versa}, the UV limit $a\to0$ should not necessarily exist. In statistical physics $a$ should not be less than the mean free path, 
or less than the lattice constant in solid state physics. In quantum field theory the 
$\lim_{a\to0}\phi_a(x)$ may have no physical meaning because the infinitely small scale 
implies an infinitely high energy due to the Heisenberg uncertainty principle.   
In any case, having the value of a coupling constant at certain scale 
$a=a_{\rm Phys}$ we ought to use the flow equation for the wavelet effective action 
$A\frac{\partial}{\partial A} \Gamma_{(A)}=\ldots$ to derive the value of coupling 
constant at other scales. 

\begin{figure}[ht]
\centering \includegraphics[width=60mm]{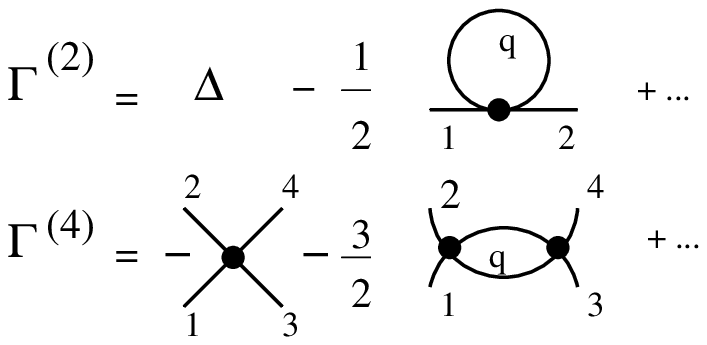}
\caption{Renormalized vertex functions. The renormalized inverse propagator and the renormalized vertex functions in $\phi^4$ theory are shown in one-loop approximation. Here and after we assume $-\lambda$ value for each vertex}
\label{v24:pic}
\end{figure}

We proceed with the $\phi^4$ theory at one loop level now. Substituting the wavelet 
transform \eqref{dwtrd} into the action \eqref{Gw} and applying the Fourier transform to 
the results we get one loop contributions, shown in Fig.~\ref{v24:pic} 

The $\Gamma^{(2)}$ and the $\Gamma^{(4)}$ one-loop contributions to the effective action 
\begin{widetext}
\begin{equation}
\Gamma_{(A)}[\phi_a] = \Gamma_{(A)}^{(0)}+\sum_{n=1}^\infty \frac{1}{n!}\Gamma_{(A)}^{(n)}(a_1,b_1,\ldots,a_n,b_n) \phi_{a_1}(b_1)\ldots \phi_{a_n}(b_n) \frac{da_1d^db_1}{a_1}\ldots 
\frac{da_nd^db_n}{a_n}
\end{equation}
\end{widetext}
are shown in Fig.~\ref{v24:pic}. They are 
\begin{equation}
C_g^2\frac{\Gamma^{(2)}_{(A)}(a_1,a_2,p)}{\tilde{g}(a_1 p) \tilde{g}(-a_2 p)} = p^2+m^2 + 
\frac{\lambda}{2} T^d_g(A) \label{g2l1},
\end{equation} 
where $d=4$ is the dimension of Euclidean space,
$A = m \min (a_1,a_2)$ is the dimensionless scale of the tadpole diagram, for the inverse 
propagator; and similarly, 
\begin{equation}
C_g^4 \frac{\Gamma^{(4)}_{(A)}}{\tilde{g}(a_1p_1)\tilde{g}(a_2p_2)\tilde{g}(a_3p_3)\tilde{g}(a_4p_4) } = \lambda -\frac{3}{2}\lambda^2 X^d_g(A) \label{g4l1}
\end{equation}
for the one loop contribution to the vertex function. 
The vertex functions $\Gamma^{(n)}_{(A)}$ obey standard functional equations, {\em e.g.,}  $G^{2}\Gamma^{(2)}=\frac{g\bar{g}}{C_g}$, which is a projection of the unity to a given scale, so that the trace is 1. 

The values of the one-loop integrals 
\begin{align}\nonumber 
T^d_g(A) &=& \frac{S_d m^{d-2}}{(2\pi)^d} \int_0^\infty f_g^2(Ax) \frac{x^{d-1}dx}{x^2+1}, \\\ 
X^d_g(A) &=& \int \frac{d^dq}{(2\pi)^d}
\frac{f^2_g(qA)f^2_g((q-s)A)}{\left[ q^2+m^2\right]\left[ (q-s)^2+m^2\right] },
\label{li1}
\end{align}
where $s\!=\!p_1\!+\!p_2, A=\min(a_1,a_2,a_3,a_4)$, 
depend on the wavelet cutoff function 
\begin{equation}
f_g(x) = \frac{1}{C_g}\int_x^\infty |\tilde{g}(a)|^2 \frac{da}{a}
\end{equation}
for the chosen wavelet $g$. 

For the simplest choice of the basic wavelet, as in \cite{Altaisky2010PRD},
$\tilde{g}_1(k) = -\imath k e^{-\frac{k^2}{2}}$,  
we get the Gaussian wavelet cutoff function $f_{g_1}(x)=e^{-x^2}$ with $C_g=\frac{1}{2}$. 
The best choice of the basic wavelet would be the apparatus function of the measuring device which defines the space of functions $\{ \phi_a(b) \}$. Working in Euclidean space we ought to make some simplifying assumptions. So, using $g_1$ as the basic wavelet we can easily calculate the beta function and the anomalous dimension of 
the effective mass in our model by taking the derivatives of equations (\ref{g2l1},\ref{g4l1}) with respect to the logarithm of scale variable $\mu = -\ln A + const$. For the 
$g_1$ wavelet  in one-loop approximation \eqref{li1} this gives the flow equations 
  \begin{align}
\frac{\partial \lambda}{\partial\mu} &=& 3\lambda^2\alpha^2 \frac{\partial X^4_1}{\partial\alpha^2} = \frac{3\lambda^2}{16\pi^2} \frac{2\alpha^2+1-e^{\alpha^2}}{\alpha^2}
e^{-2\alpha^2}, \label{b1} \\
\frac{1}{m^2} \frac{\partial m^2}{\partial\mu} &=& \frac{\lambda}{32\pi^2\alpha^2}
-\frac{\lambda}{16\pi^2} + \frac{\lambda}{16\pi^2} 2\alpha^2e^{2\alpha^2}\Ei_1(2\alpha^2),
\label{b2}
\end{align}
where $\alpha = A m$ is the dimensionless scale,  
$\Ei_1(z)=\int_1^\infty \frac{e^{-xz}}{x}dx$ is the exponential integral of the first type. 

For small $\alpha$ the equation \eqref{b1} tends to the known result. The last term in \eqref{b2} tends to zero for small $\alpha$.

To show the group composition law for the scale-dependent coupling constant $\lambda(a)$, we have to analyze a universal function $F$, that express the 
value of the coupling constant at a certain scale $a_0$ using its value at another scale $a_2$, which depends only on scale ratio and the coupling constant value:
$
\lambda(a_0) = F \left(\frac{a_0}{a_2},\lambda(a_2) \right).
$

Let us assume $a_2<a_1<a_0$ and consider the recursive relation 
\begin{align*}
F \left(\frac{a_0}{a_2},\lambda(a_2) \right) &=& F \left( \frac{a_0}{a_1}, \lambda(a_1) 
\right) \\
&=& F \left( \frac{a_0}{a_1},F\left(\frac{a_1}{a_2}, \lambda(a_2) \right) \right)
\end{align*}
in one loop approximation for the scale dependence of the coupling 
constant. Substituting 
\begin{equation}
\lambda (a_1) = \lambda(a_2) - \frac{3}{2} \lambda^2(a_2) \int_{a_2}^{a_1} d\mu(a)
\hbox{[one-loop]}
\end{equation}
into 
$$
\lambda (a_0) = \lambda(a_1) - \frac{3}{2} \lambda^2(a_1) \int_{a_1}^{a_0} d\mu(a) 
\hbox{[one-loop]}
$$
we get 
\begin{equation}
\lambda (a_0) = \lambda(a_2) - \frac{3}{2} \lambda^2(a_2) \int_{a_2}^{a_0} d\mu(a)
\hbox{[one-loop]} + O(\lambda^4),
\end{equation}
where the integrals over the measure $d\mu(a)$ were added. Therefore the value of the coupling constant depends only on its value at the reference scale and 
on the ratio of scales.

\section{Conclusions}
The method of RG has got a clear physical interpretation in the self-similarity 
picture of phase transitions. Following K.Wilson \cite{WK1974}, in the context of phase 
transitions, the RG method is usually understood as the integration over small scale degrees of freedom is such a way that the remaining large scale  degrees of freedom gain an interaction Hamiltonian different from the initial one only by scale transformation of the fields 
and coupling constants. This is accompanied by a coarse-graining procedure, performed at 
each iteration step. These iterations can  be represented by the same pair of the projection operators $H$ and $G$, that are used in discrete wavelet transform, see e.g, 
\cite{Daub10}. The coarse-graining operator $H$ is often referred to as a low-pass filter, the ''details operator'' $G$  is referred to as the high-pass filter.  Starting 
from some initial fine scale fields $s^0$ the iteration goes to coarse scales 
\begin{align*}
s^0 & \stackrel{H}{\to}      &s^1&\stackrel{H}{\to}      &s^2& \stackrel{H}{\to} & \ldots & s^n\\
    & \stackrel{G}{\searrow} &   & \stackrel{G}{\searrow}&   & \stackrel{G}{\searrow} &  & \\
    &                        &d^1&                       &d^2&               &\ldots & d^n           
\end{align*}
with the details ($d$) integrated out at each step of iteration, and the final results formulated 
on ($s^n$) space. 
Since the same functional space $\mathcal{H} = L^2(\R^d)$, or $l^2(\mathbb{Z}^d)$, is used for all steps of iteration, the coupling constant obtained in such averaging process for the rescaled fields 
$\zeta^\nu \phi(\zeta x)$ is the effective interaction constant for all scales larger than a given scale.    Surprisingly, the pair of projection operators $H$ and $G$ have been already used in 
the Ising model by G.Baker \cite{Baker1972} -- without knowing that was a Haar wavelet, -- 
who has thrown away the inter-block interactions to meet the scaling results of K.Wilson \cite{Wilson1971b} (although the integrals were analytic in their full form).

The aim of the present paper was to show that instead of a  sequence of spaces $\{ s^i\}$, which 
approximates the functional space  $\mathcal{H}$ of the considered problem, it is possible to formulate the theory completely in terms of the ''details spaces'' , i.e. on 
$\overline{\displaystyle{\cup_i d^i}}$ space, which is  dense in $\cH$. In contrast  to the nested sequence of $s^i$, the 
$d^i$ spaces are maximally uncorrelated and their sum represents the most compact record of 
information on the analyzed function \cite{Daub10,Chui1992}. It is clear from the considered model 
that the order of integration from small scales to large scales  
is not obligatory. Since the details spaces of different spaces are independent, the integration 
can be performed from large scales to small scales, adding new details at each step down to the 
prescribed finest scale $A$. This approach does not assume any divergences \cite{Altaisky2010PRD}.
The finest resolution scale $A$ is restricted only by the physics of the problem, and the existence 
of $A\to0$ limit is not required. Using the ''details spaces''  
instead of the nested set of coarse-grained spaces  we get rid of the fields renormalization and yield only the renormalization of parameters. 

The interaction terms of the form $\phi_{a_1}(b_1)\ldots\phi_{a_n}(b_n)$ in the present 
formalism describe the interaction of the field fluctuations of different scales, rather 
than the  interaction of all fluctuations up to a given scale, as in the Wilson theory. 
The physics of universality, the presence of fixed points, etc., in this representation can be 
attributed to the assumption that the scale invariance  of fundamental interactions may 
be of the same importance as their internal symmetries, and we ought to formulate the theory  on the  space   $\cup_i d^i$ of functions $\phi_a(x)$ using 
appropriate invariant measure. The difference between the resolution $a$, and the ordinary 
variables, such as $x$, $p$, etc, is that the resolution $a$ is neither a dynamical variable, nor a physical observable. (There is no Hermitian operator of scale transformations in $L^2(\R^d)$, unless it is constructed  artificially \cite{Battle1999}.) The resolution parameter $a$ can be only set for a given problem, or can be integrated over all possible values. 

\section*{Acknowledgments} 
The author is thankful to Prof. J.Polonyi for critical comments. The research was supported in part  by the RFBR Project 13-07-00409.

%

\end{document}